\documentclass[12pt]{article}
\usepackage{graphicx}
\usepackage{amssymb}

\def\vv{\vspace{-1.3ex}}

\begin{document}

\begin{center}
\thispagestyle{empty}

\vspace*{22ex}

{\Large
A lens-coupled scintillation counter in cryogenic environment
}

\vspace{7ex}
A.\,Stoykov, R.\,Scheuermann, A.\,Amato, \\
M.\,Bartkowiak, J.A.\,Konter, J.\,Rodriguez, K.\,Sedlak

\vspace{7ex}

Paul Scherrer Institut, CH-5232 Villigen PSI, Switzerland \\[1ex]

\end{center}

\vspace{13ex}
\noindent
In this work we present an elegant solution for a scintillation counter
to be integrated into a cryogenic system.
Its distinguishing feature is the absence of a continuous light guide coupling
the scintillation and the photodetector parts, operating at cryogenic
and room temperatures respectively.

The prototype detector consists of a plastic scintillator
with glued-in wavelength-shifting fiber located inside a cryostat,
a Geiger-mode Avalanche Photodiode (G-APD) outside the cryostat,
and a lens system guiding the scintillation light re-emitted by the fiber
to the G-APD through optical windows in the cryostat shields.
With a 0.8~mm diameter multiclad fiber and a 1~mm active area G-APD
the coupling efficiency of the ``lens light guide" is about 50~\%.
A reliable performance of the detector down to 3~K is demonstrated.

\clearpage
\newpage

\section{Motivation}
We have been motivated to undertake the present investigation by the necessity
to integrate a scintillation detector in the sample space of a dilution refrigerator
to be used in the High\,--\,Magnetic Field (9.5\,T) $\mu$SR spectrometer \cite{muSR,HMF}
being built at the Swiss Muon Source (S$\mu$S) \cite{LMU}
of the Paul Scherrer Institut (PSI, Villigen, Switzerland).

Figure~1 outlines the layout of the detector system of the spectrometer.
The high time resolution counters Mt and Pt measure the time interval between
a muon entering the sample and its decay positron being emitted.
To ensure the required performance, these detectors are operated at  room temperature,
their prototypes are described in \cite{Stoykov_PhysB404-990}.
Inside the cryostat, in the vicinity of the sample, the so-called muon veto Mv
and positron validation Pv detectors are located.
Mv serves the purpose of additionally collimating the incoming muon beam:
only those muons which pass through the central hole in this counter will be accepted
by the data acquisition system.
Pv works in coincidence with Pt: it confirms that a decay positron detected by Pt
has originated from a small inner volume surrounded by Pv --
this is expected to reduce the intensity of random Mt-Pt coincidences
contributing to the uncorrelated background in the $\mu$SR spectra.
The Pv detector has a cylindrical shape and is divided into 4 segments
of independent scintillation counters to provide additional information
about the direction at which a decay positron has been emitted.
Due to space limitations we can not afford more than 5~mm thickness for a Pv counter
(taking into account spiraling of the decay positrons in high magnetic fields,
the diameter of the Pt detector ring, and accordingly the cryostat diameter,
should be kept as small as possible \cite{Kamil_HMF-simulations}).
Pv represents the most critical part in terms of ensuring the required performance:
relativistic decay positrons passing through this counter deposit a factor of 3 less energy
compared to 29~MeV/$c$ muons stopping in Mv.

An ideal choice for Pv (and Mv) would be a detector
consisting of a plastic scintillator,
a wavelength-shifting (WLS) fiber to collect and transport the scintillation light, and
a Geiger-mode Avalanche Photodiode \cite{G-APD} as a photosensor.
Such type of detector has been shown to provide the necessary performance,
also in high magnetic fields, and compactness \cite{newALC}.
The problem, however, is how to integrate such a system into the cryostat.

In the construction of the cryostat we distinguish two major parts (see Figure~1):
the cryostat body, where the vital components necessary for cooling are located,
and the removable tail, including only the radiation shields and the outer vacuum chamber
(removable tail part allows for the sample change).
A straightforward detector solution suggests a continuous (presumably fiber) light guide
going to the photosensor through the body of the cryostat.
This, however, interferes with the foreseen cryostat design and is difficult
to implement without compromises.
An alternative solution is to collect the light from the fiber by an optical lens system,
to make a parallel light beam, reflect it by 90 degrees at a mirror,
and transport it through optical windows in the cryostat shields to the photosensor.
This approach can be realized by only modifying the cryostat tail section,
keeping the cryostat body untouched.

We point out, that in scintillation counting,
where each individual particle has to be detected
with, ideally, 100\,\% efficiency and often with preserved timing and energy information
in the signal, an optical lens system is practically not used at all:
in the majority of the cases a continuous light guide will deliver more light and is
much simpler to realize.
To the best of our knowledge,
the only case where a lens system was used in the construction
of a scintillation counter is reported in \cite{Kerek98}.
The present work gives another example of an efficient lens-coupled
scintillation detector: although by idea similar to \cite{Kerek98},
it is completely original work concerning the technical details.

\section{Lens light guide}
An optimal design of the ``lens light guide" (LLG),
complying with our space requirements, was deduced experimentally.
We used a UV-lamp to excite a wavelength shifting fiber
and tried different combinations of lenses
to collect the emitted light with maximum efficiency.
The light beam at different points of our system
was visualized using semitransparent diffuse screens.
All optical components were obtained from Edmund Optics Inc. \cite{EdmundOptics}.

Figure~2 gives the LLG details:
the transmitting lens (T-Lens) collects the light from the fiber;
the resulting slightly diverging light beam is reflected by 90 degrees,
and sent to the receiving lens (R-Lens), that focusses the light onto a G-APD.
The T-Lens, to be mounted inside a cryostat, should be as compact, as possible:
as a compromise between the compactness and the ease of handling
lenses with 6\,mm diameter were used.
The maximum emission angle of the multiclad WLS fiber (Bicron BCF-92)
is $\theta = 48^\circ$, which corresponds to a numerical aperture
$\emph{NA} = \sin\theta = 0.74$. To collect all this light,
the object space numerical aperture of the T-Lens should be equal or larger.
This is ensured by using, as a first element, a large numerical aperture
aspherical lens at zero distance from the fiber;
this lens is followed by two plano-convex lenses
reducing the divergence of the light beam.
The R-Lens was built in two versions,
depending on the active area of the photosensor to be used:
the R-Lens-3 is used with a 3\,x\,3~mm$^2$ Hamamatsu MPPC S10362-33-050C \cite{Hamamatsu},
the stronger R-Lens-1 allows to focus the light onto a smaller device --
$\oslash 1.1$~mm Photonique SSPM 0810G1MM \cite{Photonique}.
The R-Lens is positioned outside the cryostat, without strict space requirements.
Depending on the R-Lens version we distinguish two versions of the lens light guide:
LLG-3 (T-Lens + R-Lens-3) and LLG-1 (T-Lens + R-Lens-1) used with
3\,x\,3\,mm$^2$ and $\oslash 1.1$\,mm G-APDs, respectively.

The coupling efficiency \emph{CE} of LLG-3 was obtained in measurements with
a $^{90}$Sr radioactive source. The 3\,x\,3~mm$^2$ active area G-APD was coupled to
the fiber either directly (no optical grease was used), or via the lens light guide.
Comparing the signal amplitudes (their mean values) in both cases, see Figure~3,
we obtain the \emph{CE} of LLG-3 as $\approx 75$\,\%
(the missing 25\,\% are, presumably, reflection losses).
Direct coupling of $\oslash 1.1$~mm G-APD to the fiber was not possible,
since the last two elements of R-Lens-1 were glued onto it (reason discussed below).
Consequently, the coupling efficiency of LLG-1 was determined in relative measurements.
We used the UV\,-\,lamp to excite the fiber and measured,
at a small bias voltage ensuring unity gain,
the corresponding G-APD photocurrent $I_{\rm ph}$.
The measurement was performed in three steps by using the 3x3~mm$^2$ G-APD:
1)~$I_{\rm ph}$ was measured with R-Lens-3;\
2)~R-Lens-3 was replaced by R-Lens-1 -- the photocurrent was the same as in
the previous measurement, since the light spot at the G-APD position in both cases
is smaller than 3~mm;
3)~measurements with the R-Lens-1 were continued,
but the G-APD active area was restricted to $\oslash 1.1$~mm by placing
in front of it a thin non-transparent disc with a central hole.
The photocurrent was now at 73\,\% of its maximum value, measured before.
By combining the 75\,\% absolute \emph{CE} of LLG-3
with 73\,\% light collection of R-Lens-1 to $\oslash 1.1$~mm photosensor,
we estimate the absolute coupling efficiency of LLG-1 as $\approx 55$\,\%.

Taking into account that the T-Lens will be mounted inside a cryostat,
it will be moving relative to the R-Lens depending on the operation temperature
due to the thermal contraction of the cryostat materials.
We point out, that in our case with the dilution refrigerator,
the scintillation part of the detector and, accordingly,
the T-Lens, will be mounted on the 1\,K radiation shield
and hence always be at a fixed position during the cryostat operation.
Nevertheless, even in this case one should know what displacement
of the light beam at the entrance of the R-Lens can be tolerated.
Up to 8 independent readout channels will be used in the real system,
one for each detector segment of Pv and Mv.
A slight error in mounting of the T-Lens mirror might lead to
a significant displacement of the beam spot at the entrance of the R-Lens,
and this displacement can not be compensated simultaneously in all the channels.

The dependence of the G-APD photocurrent on the displacement ($\Delta {\rm x}$)
of the T-Lens relative to the R-Lens is shown in Figure~4.
The signal amplitude at $\Delta {\rm x} = \pm 2$~mm is 90\,\% and 70\,\%
of its maximum value with R-Lens-3 and R-Lens-1, respectively.
Figure~5 shows the light spot position in the above two measurements
at the entrance of the R-Lens and at the G-APD position:
2~mm displacement of the T-Lens leads to $\sim 0.5$\,mm displacement of the image
in both versions of the receiving lens:
the displacement reduction factor is therefore $\sim 4$.

Table~1 summarizes the characteristics of the two versions of the detector
differing in the LLGs and the G-APDs.
Thanks to the fact that the smaller active area G-APD used with LLG-1 has
a factor of 2 lower noise level compared to the larger one used with LLG-3,
the signal-to-noise ratio in the first case is a factor of 1.4 better,
although the amount of collected light is smaller.
Considering the insensitivity of the LLG-1
to the relative displacement of its transmitting and receiving lenses as acceptable,
further tests in a cryogenic system were performed using this version of the lens
light guide.

\begin{table}[htb]
\begin{center}
\caption{
Characteristics of the two detector versions
using different lens light guides and different photosensors.
By the photosensor noise $A_{\rm noise}$ we understand a minimum threshold value
to suppress the thermal noise count rate to less than $\sim 10$ counts per second.
Due to cross-talk this value is several times the amplitude $A_{\rm 1e}$ of the signals
from the breakdown of single G-APD cells.
The signal-to-noise ratio $S/N$ is defined as $A / A_{\rm noise}$,
where $A$ is the mean signal amplitude.
Operating conditions of the G-APDs are specified in the caption of Figure~3.
}
\vspace{2ex}
\begin{tabular}{|l|c|c|}
\hline
 & \multicolumn{2}{c|}{Detector version} \\
\cline{2-3}
 & LLG-3 + MPPC & LLG-1 + SSPM \\
\hline
\underline{Lens Light Guide} & & \\
\ \ -~coupling efficiency & 75\,\% & 55\,\% \\
\ \ -~light spot size at photosensor & $\sim 2$\,mm & $\sim 1$\,mm \\
T-Lens vs. R-Lens  displacement ($\Delta {\rm x}$): & & \\
\ \ -~displacement reduction factor & $\sim 4$ & $\sim 4$ \\
\ \ -~relative signal amplitude at $\Delta {\rm x}= 2$\,mm & 0.9 & 0.7 \\
\underline{Photosensor} & & \\
\ \ -~active area & 3\,x\,3\,mm$^2$ & $\oslash 1.1$\,mm \\
\ \ -~noise level ($A_{\rm noise} / A_{\rm 1e}$) & $\sim 6.7$ & $\sim 3.3$ \\
\underline{Detector as a whole} & & \\
\ \ -~$S/N$ with $^{90}$Sr\,-\,source & 5.7 & 8 \\
\ \ -~$S/N$ with positron beam        & --  & 11 \\
\hline
\end{tabular}
\vspace{-2ex}
\end{center}
\end{table}

\newpage
\section{Cryogenic tests}
For cryogenic tests we adapted the existing cryostat
of the ALC $\mu$SR spectrometer \cite{ALC}:
both the outer vacuum chamber and the radiation shield were split into two parts --
a long ($\sim 70$~cm) base and a short ($\sim 15$~cm) removable front tail
housing the detector (see Figure~6).
The tests were performed with a $\sim 29$~MeV/$c$ positron beam
at the $\pi$E3 beamline of S$\mu$S.

Figure~7 shows the details of the detector built-in into the cryostat tail:
the inner module includes the scintillators, WLS fibers, and T-Lenses,
while the R-Lens assemblies with G-APDs are mounted on the outer vacuum chamber.
The construction allows integrating up to 8 scintillation counters.
The WLS fiber is glued into the groove in the scintillator and extends towards the T-Lens.
In front of the scintillator the fiber continues to be glued into a ring
made of an UV\,-\,transparent plastic (Bicron BC-800).
This is done to allow the UV-light (370~nm) shining through
the sapphire entrance window during the alignment procedure to be collected by the fiber
(this light can not penetrate the scintillator because of Teflon wrapping used to improve
the light collection by diffuse reflection).
At the T-Lens side all the fibers (there can be up to 8 of them)
are glued into a plexiglass ring, their ends are polished.
This ring, the fibers, and the scintillators form a stand-alone module, onto which
the T-Lens is mounted. The T-Lens assembly attached, the whole inner detector module is fixed
on a cylindrical Al\,-\,shield, which is put into thermal contact with
the cryostat cold finger via an adapter piece
(a thread connection allows adjusting the detector position relative to the R-Lens).
On the way towards the R-Lens the light passes through three windows,
two in the shields and one in the outer vacuum chamber.
The windows in the shields are made of BK-7 glass,
which effectively cuts the room temperature radiation.
Since the incidence angle of the light rays on the windows is small,
the reflection losses are negligible (all the windows have antireflection coatings).
For mounting the elements of the R-Lens, a standard C-Mount system is used.
The last two elements in the R-Lens-1 (the PCX\,6x6 lens and the spacing window)
are glued to the G-APD. In this particular G-APD the cross-talk is partially determined
by the internal reflection of light emitted during the avalanche
at the boarder between the epoxy covering and the air.
By gluing a piece of glass onto the front face of the G-APD  the amount
of reflected light is reduced, which leads to the noise level reduction by a factor of 2.
The G-APD is connected to an amplifier (gain $\sim 10$, bandwidth $\sim 100$~MHz)
via a single coaxial cable for the bias voltage and the output pulse.
The amplifier scheme is based on the one described in \cite{JINST06}.

We performed three cooling cycles of our system
(from 300\,K to 3\,K -- the base temperature of our cryostat).
For each new cycle the cooling and heating rates were increased.
Figure~8 shows the temperature history in the last, shortest, cycle:
the cooling rate was not less than 3~K/min and the heating rate reached even 10~K/min.
No problems with the mechanical stability of the system were encountered.

At cooling from 300\,K to 3\,K the contraction of the sample stick,
onto which the inner detector module is mounted, is about 2.5\,mm.
Accordingly, the signal amplitude at a certain temperature will depend
on the initial alignment of the T-Lens in respect to the R-Lens.
During the last cooling cycle the alignment was done in such a way
to maximize the signal at low temperatures.
Figures~9~-~11 show, respectively:
dependence of the signal amplitude (mean value) vs. temperature;
light spot at the entrance of the R-Lens and at the G-APD position;
detector signals at different temperatures.
During all cryogenic tests the performance of the detector has been stable 
with the signal-to-noise ratio well exceeding the acceptable level
in the whole temperature range.

\section*{Summary}
In this work we built a lens-coupled scintillation counter
and demonstrated its reliable performance both at room and cryogenic temperatures.

We point out that the use of lenses offers an elegant solution
for a scintillation detector to be integrated into a cryogenic system.
To the best of our knowledge, such idea has never been realized before.

\section*{Acknowledgements}
We express our gratitude to Urs Greuter for developing the amplifier,
and Matthias Elender for his valuable help in designing and assembling
the test cryogenic system.

\clearpage
\newpage
\begin{figure}[p]
\begin{center}
\includegraphics*[width=16cm]{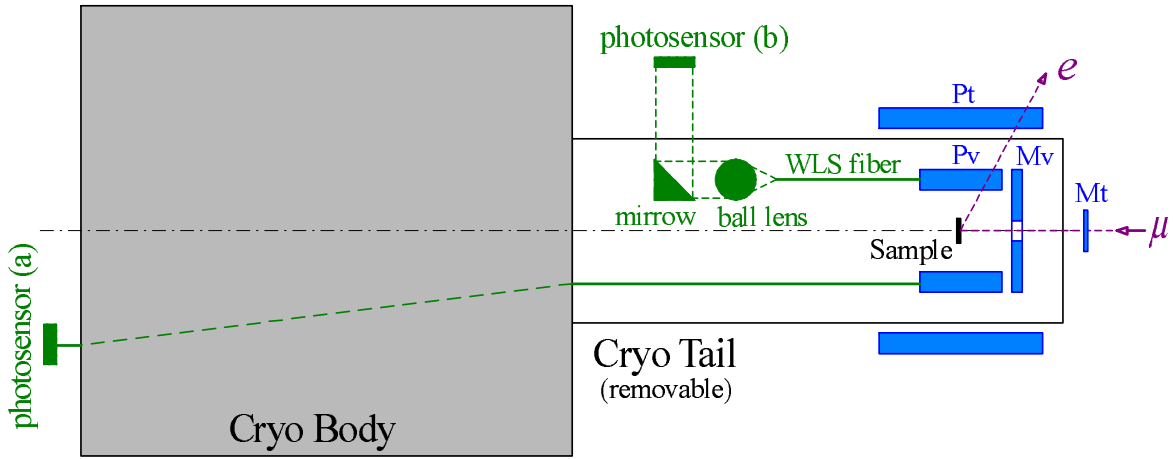}\\[1ex]
\caption{
The sample environment and the detector layout
of the High\,--\,Magnetic Field (9.5\,T) $\mu$SR instrument.
The sample is mounted on the cold finger
of the cryostat (dilution refrigerator).
The radiation shields and the outer vacuum chamber in the tail part of the cryostat
are removable to allow for the sample change.
Mt and Pt are the high time resolution muon and positron counters
operated at room temperature.
Mv and Pv are muon veto and positron validation counters:
the scintillators with embedded wavelength shifting fibers
are located in the vicinity of the sample on the 1\,K radiation shield of the cryostat,
the photosensors are at room temperature.
Two envisaged optical coupling schemes in Mv and Pv are:
a)~a continuous, presumably fiber, light guide;
b)~a discontinuous lens light guide:
the fiber is terminated shortly after the scintillator;
further on the light is transported by a lens system through
the windows in the cryostat shields.
Compared to (a), the scheme (b) requires minimum adaptation of the cryostat design
to integrate the detector.
}
\end{center}
\end{figure}

\clearpage
\newpage
\begin{figure}[p]
\begin{center}
\includegraphics*[width=15cm]{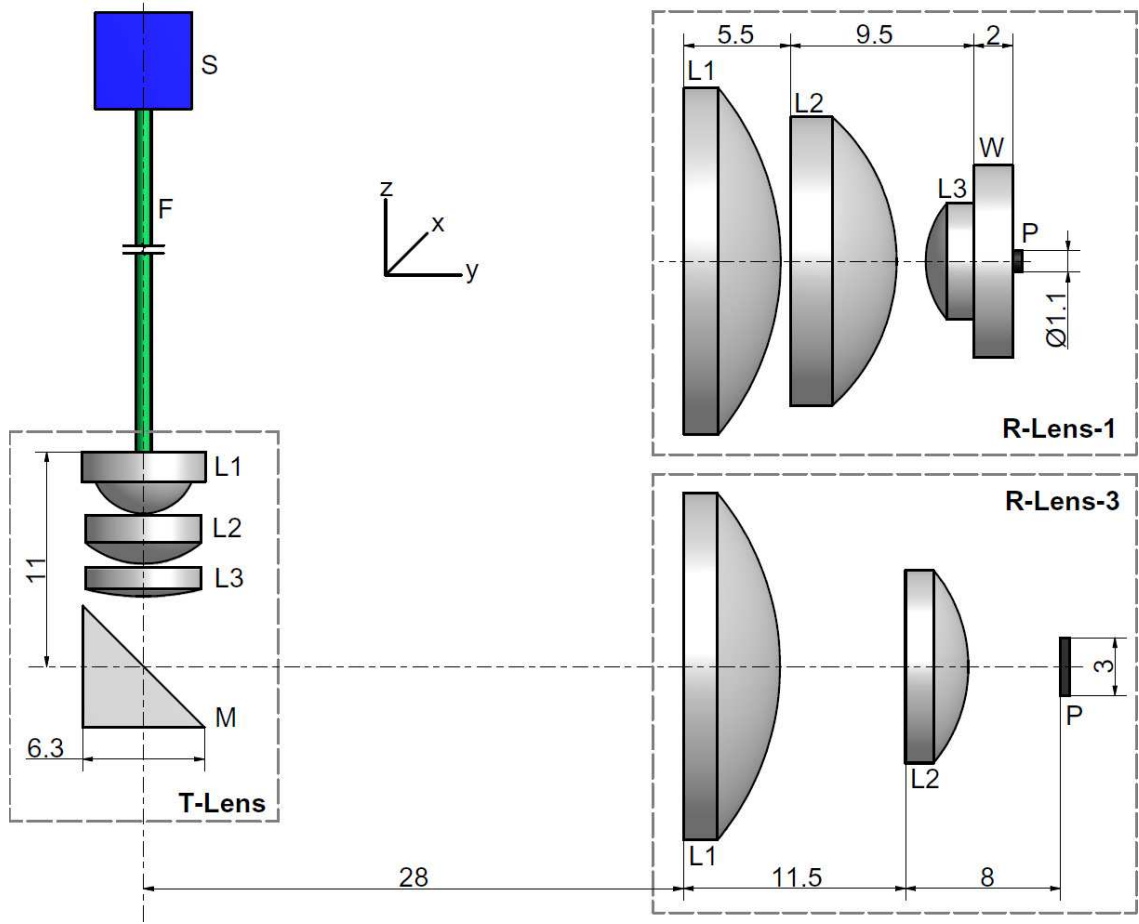}\\[1ex]
\caption{
Scheme of light collection from the scintillator (S), via the WLS fiber (F) and
the lens light guide (LLG), to the photosensor (P):
S~--~5\,mm thick EJ-204 plastic scintillator from Eljen \cite{Eljen} with a 1\,mm deep
groove to accommodate the fiber; \
F~--~$\oslash 0.8$\,mm multiclad BCF-92 WLS fiber from Bicron \cite{Bicron} glued
into the scintillator using an Eljen EJ-500 optical cement; \
P~--~photosensor.
The transmitting T-Lens collects the light from the fiber
and sends it to the receiving R-Lens, to be focused onto the G-APD.
Depending on the active area of the photosensor,
there are two versions of the R-Lens.
The used optical components are listed below.
\underline{T-Lens}:\
L1~--~large numerical aperture aspherical lens (\emph{NA}\,=\,0.62),\
L2~--~plano-convex lens PCX\,6x6,\
L3~--~PCX\,6x24,\
M~--~right-angle mirror;\quad
\underline{R-Lens-3}:\
L1~--~PCX\,18x18,\
L2~--~PCX\,10x10;\quad
\underline{R-Lens-1}:\
L1~--~PCX\,18x18,\
L2~--~PCX\,15x15,\
L3~--~PCX\,6x6,\
W~--~window $\oslash 10$\,x\,2~mm used as a spacer between the G-APD
and the L3 lens. The window is glued to the G-APD with EJ-500 epoxy, and to the lens
with a UV-light cured optical adhesive NOA-65 from Edmund Optics.
The reason for gluing the window onto the G-APD is discussed in the text.
}
\end{center}
\end{figure}

\clearpage
\newpage
\begin{figure}[p]
\begin{center}
\includegraphics*[width=7cm]{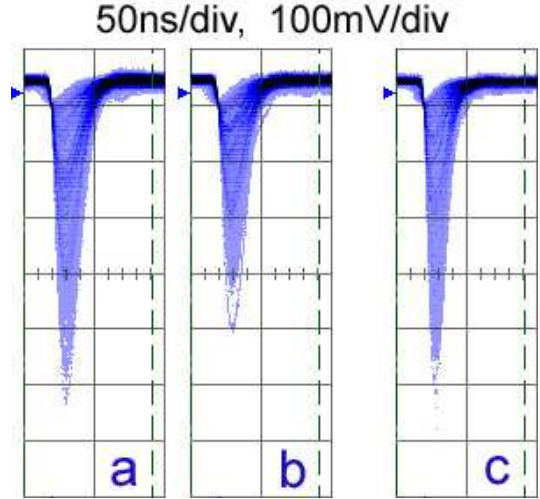}\\[1ex]
\caption{
Detector signals obtained in table tests with a $^{90}$Sr radioactive source:
a)~MPPC directly coupled to the fiber (mean signal amplitude $A = 213$~mV);
b)~MPPC with LLG-3 ($A = 160$~mV);
c)~SSPM with LLG-1 ($A = 211$~mV).
Operating conditions of the G-APDs were:
\underline{MPPC}: bias voltage $U_{\rm bias} = 70.3$~V, dark current $I_0 = 2.2~\mu$A,
amplitude of the 1e-signals from the breakdown of single G-APD cells
$A_{\rm 1e} = 4.2$~mV, maximum noise amplitude $A_{\rm noise} \approx 28$~mV;
\underline{SSPM}: $U_{\rm bias} = 29.4$~V, $I_0 = 1.3~\mu$A, $A_{\rm 1e} = 7.8$~mV,
$A_{\rm noise} \approx 26$~mV.
The used amplifier (bandwidth $\sim 100$~MHz, gain $\sim 10$) is based
on the one described in \cite{JINST06}.
}
\end{center}
\end{figure}

\clearpage
\newpage
\begin{figure}[t]
\begin{center}
\includegraphics*[width=13cm]{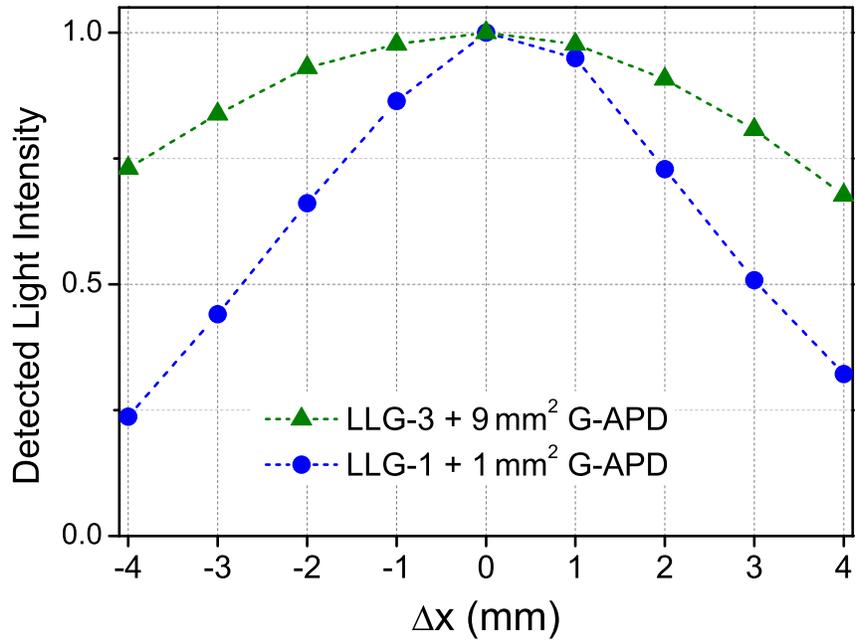}\\[-3ex]
\caption{
LLG-3 and LLG-1 coupling efficiency (relative to maximum)
as a function of the displacement ($\Delta {\rm x}$)
of the T-Lens relative to R-Lens.
}
\end{center}
\end{figure}

\begin{figure}[b]
\begin{center}
\includegraphics*[width=11cm]{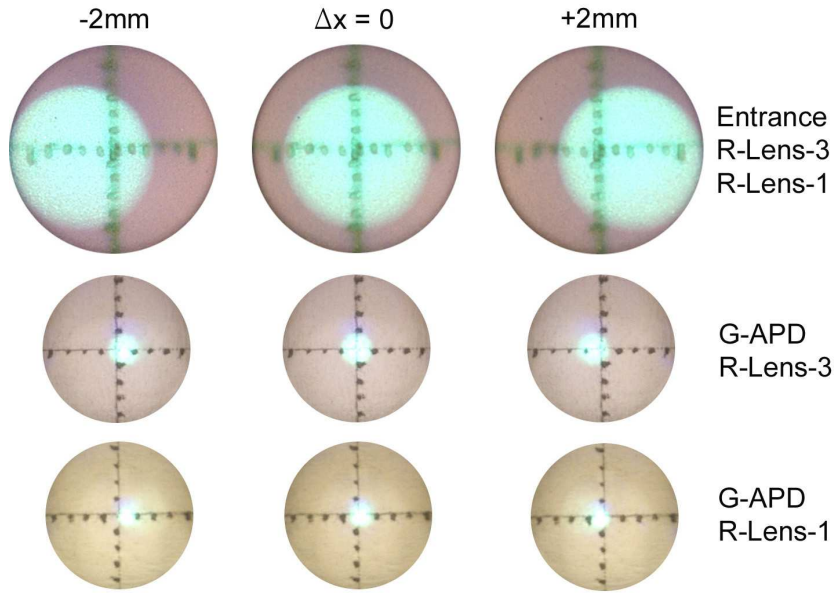}\\[1ex]
\caption{
Light spot at the entrance of the R-Lens and at the G-APD position
at different displacements ($\Delta {\rm x}$) of the T-Lens.
To obtain these images semitransparent diffuse screens $\oslash 14$~mm and $\oslash 9$~mm
were used (the scaling on each screen is 1~mm).
}
\end{center}
\end{figure}

\clearpage
\newpage
\begin{figure}[p]
\begin{center}
\includegraphics*[width=13cm]{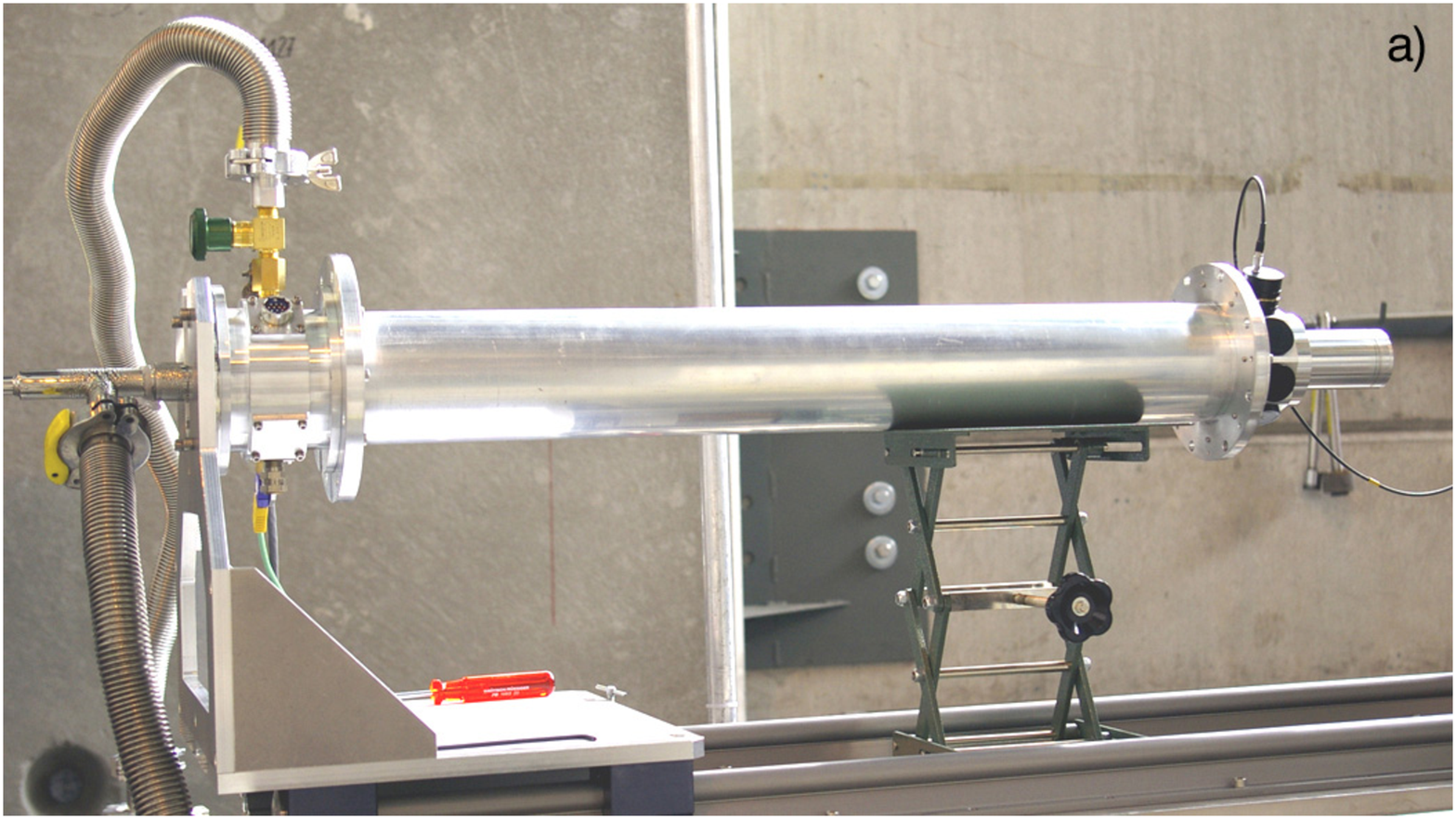}\\[3ex]
\includegraphics*[width=9cm]{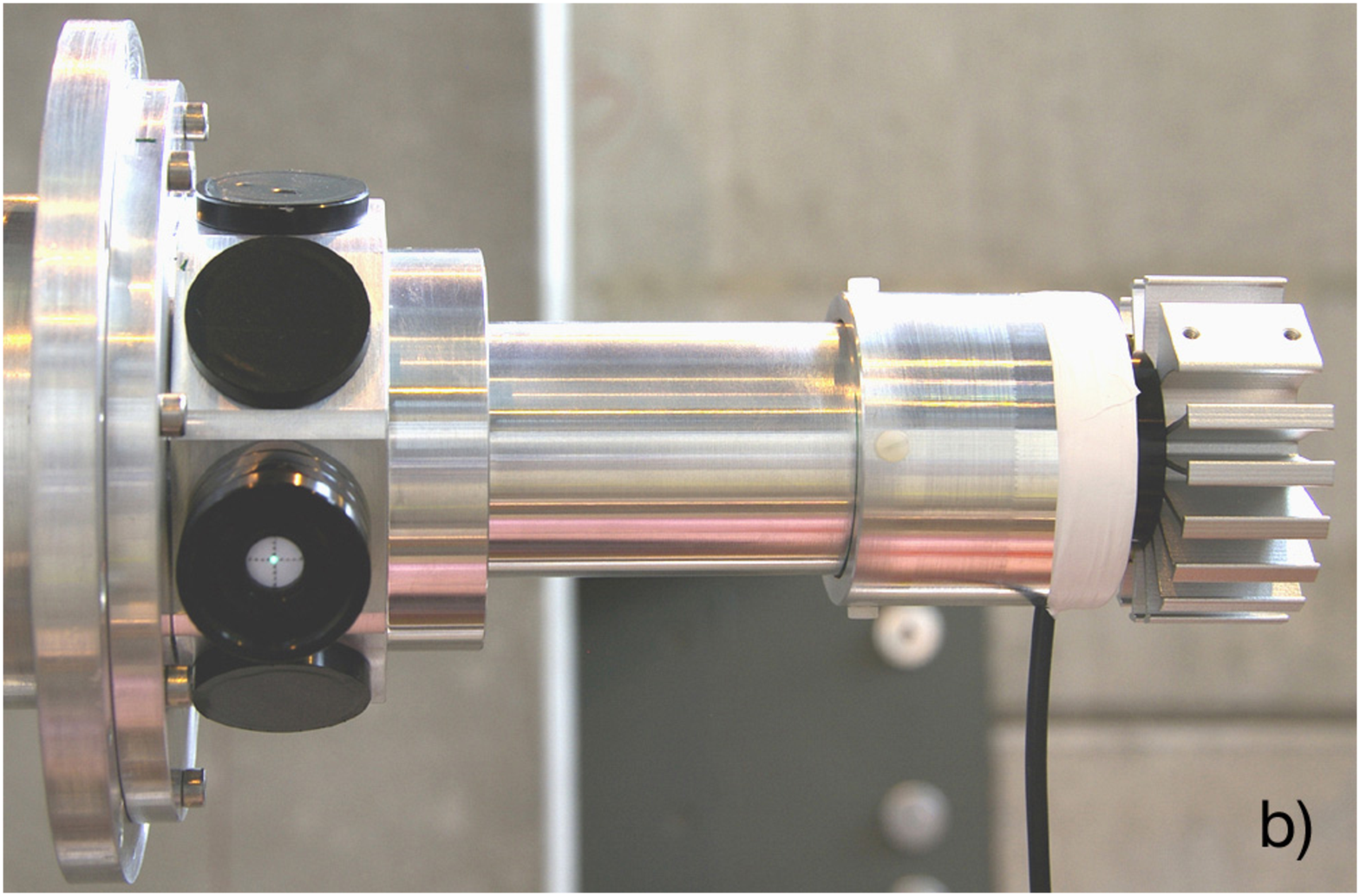}\hspace{0.5cm}
\includegraphics*[width=3.5cm]{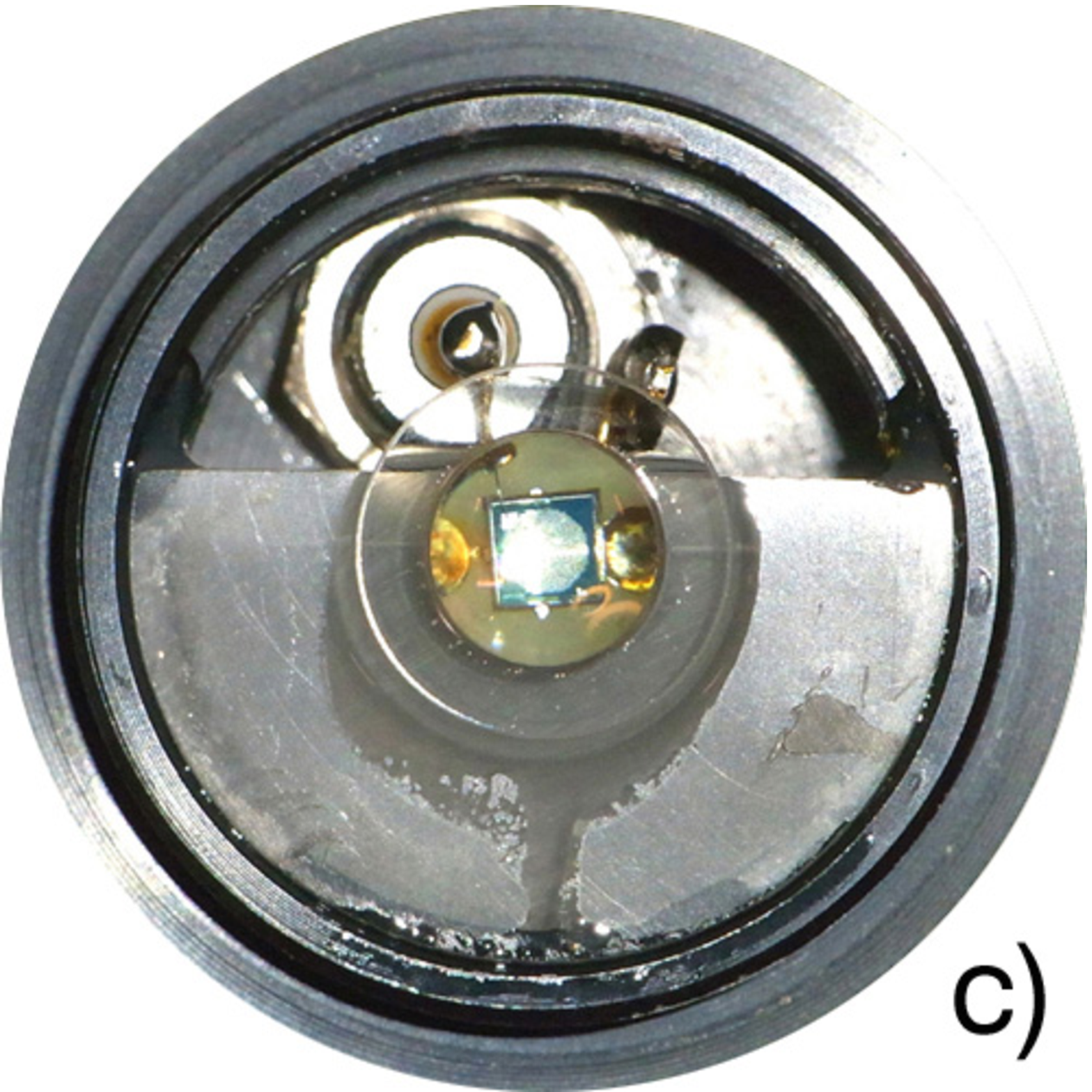}\\[2ex]
\caption{
Components of the test cryogenic system:
a)~He-flow cold-finger cryostat (base $T \approx 3$~K), consisting of
a long ($\sim 70$~cm) base and a short ($\sim 15$~cm) front tail housing the detector;
b)~the tail part with mounted UV-lamp to check the alignment of the detector.
One can see the light spot at the G-APD position in one of the readout channels.
During tests with the positron beam, the lamp will be removed and
replaced with a light-tight cap;
c)~G-APD module (C-Mount compatible) with $\oslash 1.1$\,mm G-APD:
$\oslash 10 {\rm x} 2$~mm spacing glass and PCX\,6x6 lens are glued on top.
}
\end{center}
\end{figure}

\clearpage
\newpage
\begin{figure}[p]
\begin{center}
\includegraphics*[width=16cm]{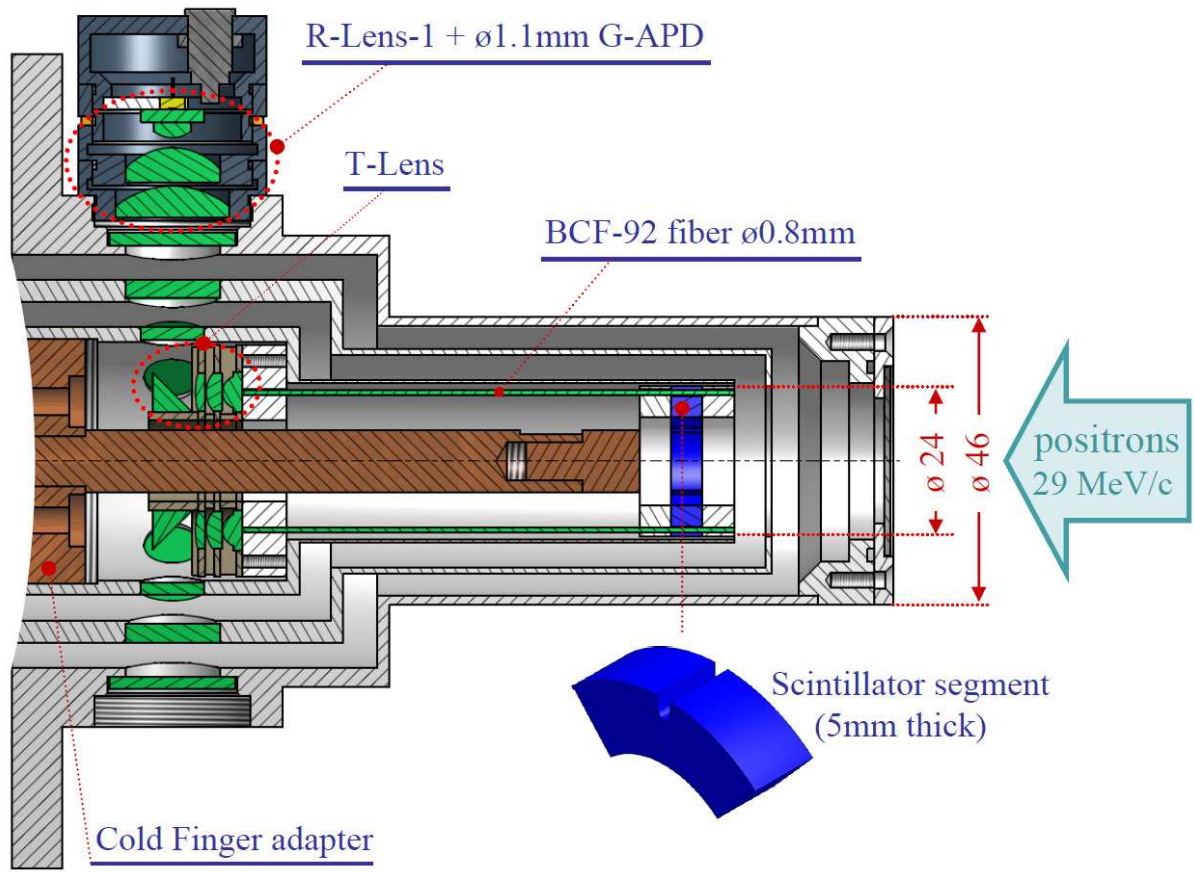}\\[2ex]
\caption{
Design view of the cryostat tail part with build-in detector.
The construction allows having up to 8 independent detector channels.
The fibers are glued into the scintillators and the plexiglass supporting rings
using EJ-500 epoxy.
The T-Lens is assembled by stacking together component and spacing layers.
Each component layer holds up to 8 identical optical elements arranged onto a circle:
the lenses are glued into the corresponding holes using
NOA-65 UV\,-\,cured optical adhesive, the mirrors are fixed with the Stycast epoxy.
}
\end{center}
\end{figure}

\clearpage
\newpage
\begin{figure}[t]
\begin{center}
\includegraphics*[width=13cm]{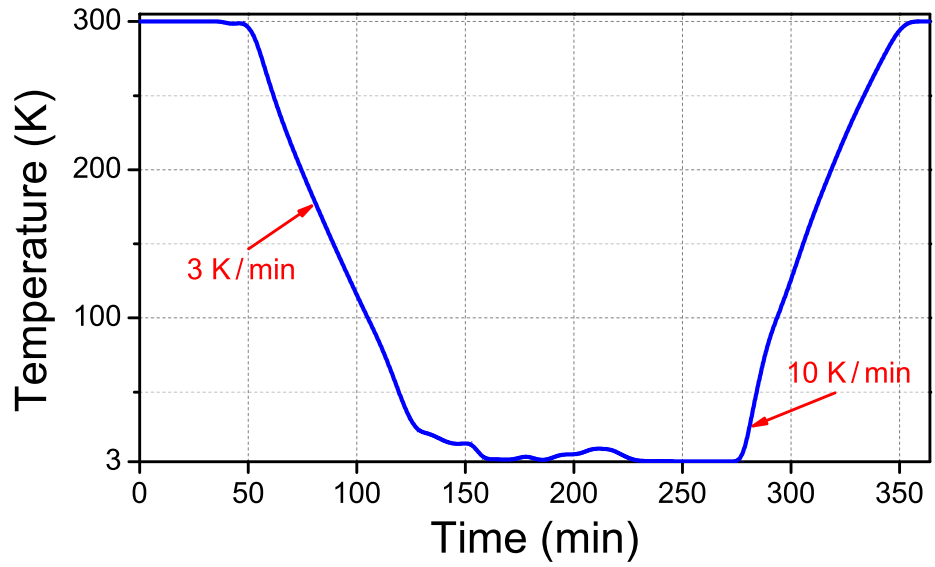}
\caption{
Cryogenic tests: temperature history in the shortest cooling cycle.
}
\end{center}
\end{figure}

\begin{figure}[b]
\begin{center}
\includegraphics*[width=13cm]{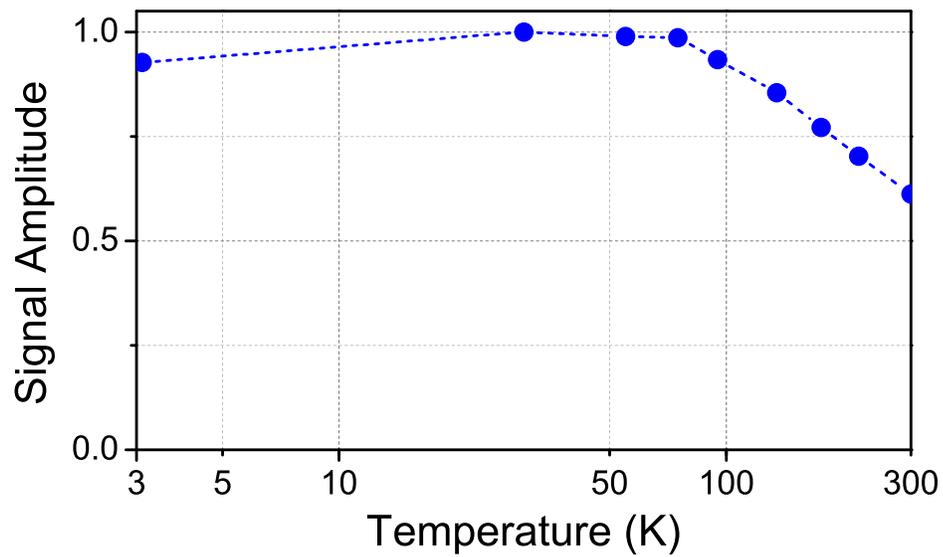}
\caption{
Amplitude of the detector signals as a function of temperature.
The change is conditioned by the displacement of the inner detector module
relative to the R-Lens, caused by the thermal contraction of the sample stick.
}
\end{center}
\end{figure}

\clearpage
\newpage
\begin{figure}[t]
\begin{center}
\includegraphics*[width=9cm]{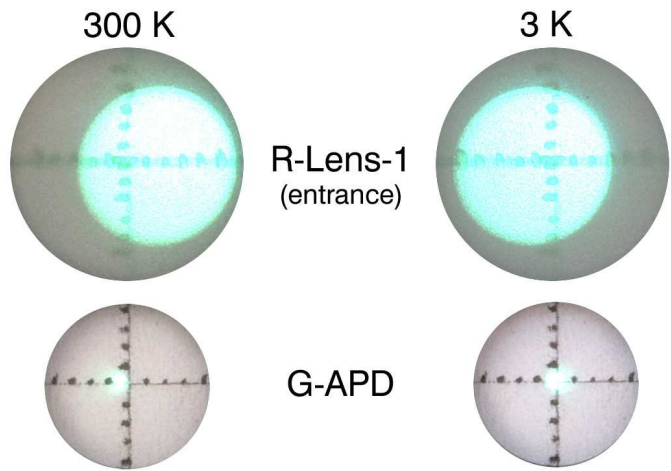}
\caption{
Light spot at the entrance of the R-Lens and at the G-APD position
at the room and the base temperatures.
To obtain these images semitransparent diffuse screens $\oslash 14$~mm and $\oslash 9$~mm
were used (the scaling on each screen is 1~mm).
}
\end{center}
\end{figure}

\begin{figure}[b]
\begin{center}
\includegraphics*[width=10cm]{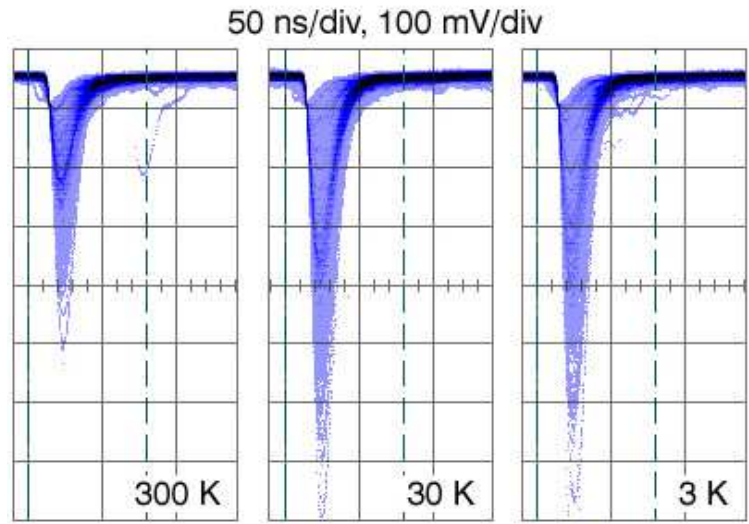}
\caption{
Detector signals at different temperatures:
\underline{300\,K}:~$A = 177$~mV, $S/N = 7$;\quad
\underline{30\,K}:~$A = 289$~mV, $S/N = 11$;\quad
\underline{3\,K}:~$A = 268$~mV, $S/N = 10$.
G-APD noise level $A_{\rm noise} \approx 26$~mV.
}
\end{center}
\end{figure}

\end{document}